%% file: EnglishCorrection.tex
\documentclass[longauth]{aa}
\usepackage[varg]{txfonts}

\begin{document}

\title{Detection of bridge emission above 50 GeV\\ from the Crab pulsar with the MAGIC telescopes}



\input{authors_last_AA_140210}


\date{Submitted 18 February 2014, Accepted 24 April 2014}

\abstract {The Crab pulsar is the only astronomical pulsed source detected at very high energy (VHE, E>100GeV) {gamma rays}. The emission mechanism of VHE pulsation is not yet fully understood, although several theoretical models have been proposed.} {In order to test {new models}, we measured the light curve and the spectra of the Crab pulsar with high precision by means of deep observations.}
 {We analyzed 135 hours of selected MAGIC data taken between 2009 and 2013 in stereoscopic mode. In order to discuss the spectral shape in connection with lower energies, {5.5} years of {\it Fermi}-LAT data were also analyzed.}
 {The known two pulses per period were detected with a significance of $8.0~\sigma$ and $12.6~\sigma$. In addition, significant emission was found between the two pulses with $6.2~\sigma$.} 
 {We discovered the bridge emission above 50 GeV between the two main pulses.  This emission can not be explained with the existing theories. These data can be used for testing new theoretical models.}


\maketitle 

\titlerunning{Discovery of bridge emission from the Crab pulsar above 50 GeV}

\section{Introduction}

The \object{Crab pulsar} and the surrounding \object{Crab nebula} are the remnant of the supernova of {AD} 1054. 
 It is one of the youngest pulsars known and its spin down luminosity ($4.6 \times 10^{38}$ erg/s) is the highest among Galactic neutron stars.  A remarkable feature of the Crab pulsar is that it is visible at all wavelengths, from radio ($10^{-5}$ eV) to VHE {gamma rays} ($> 10^{11}$ eV). To date, this pulsar is the only one for which pulsed emission has been detected above 100 GeV. 
  
 Gamma-ray pulsation from the Crab pulsar up to $\sim10$~GeV had been known since the 1990s from EGRET observations \citep{1993ApJ...409..697N}. 
 In 2008, pulsations were found by the MAGIC telescope at energies above 25~GeV \citep{2008Sci...322.1221A}. This result suggested that the emission originates in the outer magnetosphere. 
 The simplest curvature radiation scenario in the outer magnetosphere predicts an exponential cutoff in the energy spectrum at GeV energies \citep[e.g.,][]{2004ApJ...606.1143M,2006MNRAS.366.1310T, 2008ApJ...676..562T}.  {\it Fermi}-LAT observations from 100 MeV to a few tens of GeV, which started in August 2008, showed a clear break in the spectrum at $\sim 6$~GeV \citep{2010ApJ...708.1254A} supporting this scenario.
 A few years later, however, MAGIC and VERITAS \citep{2011ApJ...742...43A, 2012A&A...540A..69A,2011Sci...334...69V}  found that the energy spectrum of the Crab pulsar extends up to 400~GeV following a power law. The emission above 100~GeV is difficult to explain {only} with the curvature radiation,  and additional or different  emission mechanisms {are} required. Several new models were recently proposed 
  {to} explain the energy spectrum of the Crab pulsar {\citep[e.g.,][]{2011ApJ...742...43A, 2012Natur.482..507A}}. 
  
  Here we present new results {from} the continuing monitoring of the Crab pulsar with the MAGIC telescopes that will help to constrain any model for the emission. In order to discuss the Crab pulsar spectra at energies lower than those accessible to MAGIC, {\it Fermi}-LAT data were also analyzed.
  
 \section{Instruments, data sets, and analysis methods}
 \subsection{The MAGIC Telescopes}
  The MAGIC telescopes are two Imaging Atmospheric Cherenkov Telescopes located on the island of La Palma (Spain) at 2200~m above {sea} level. Both telescopes consist of a 17~m diameter reflector and a fast imaging camera with a field of view of $3.5\degr$. The trigger threshold for regular observations at zenith angles below 35\degr~is around 50~GeV and the sensitivity above 290~GeV (in 50~h) is 0.8\% of the Crab nebula flux with an angular resolution better than 0.07$\degr$ \citep{2012APh....35..435A}.  The first telescope started operation in 2004, while the second one became operational in 2009.
  
 For this study we used 135 hours of data taken at zenith angles below 35\degr~during optimal technical and weather conditions between September 2009 and April 2013. 
 Standard MAGIC analysis, as described in \citet{2009arXiv0907.0943M} and \citet{2012APh....35..435A}, was applied to the data.
 The conversion from event arrival times to pulsar rotational phases used {\sf Tempo2} software \citep{2006MNRAS.369..655H} and a dedicated package inside MARS \citep{MarcosThesis}. The spin parameters of the Crab pulsar were taken from the monthly reports of the Jodrell Bank Radio telescope\footnote{http://www.jb.man.ac.uk/\~{}pulsar/crab.html} \citep{1993MNRAS.265.1003L}.

 \subsection{{\it Fermi}-LAT }
 The Large Area Telescope (LAT) is a pair conversion gamma-ray detector on board the {\it Fermi} satellite \citep{2009ApJ...697.1071A}. It can {detect} high{-}energy {gamma rays} from 20 MeV to more than 300 GeV. It has been operational since August 2008 and all the collected data are publicly available. In this work, we have used {5.5} years of 
 Pass 7 {reprocessed} data\footnote{{http:\slash \slash fermi.gsfc.nasa.gov\slash ssc\slash data\slash analysis\slash documentation\slash Pass7REP\_usage.html}}
 from 2008 August 4 to {2014 January 31.} {The region of interest was chosen to be 30\degr~around the Crab pulsar. }

%
  

Along with the public data, the LAT team provides the
corresponding analysis software and 
instrument response functions (IRF) designed for the analysis of that particular dataset. We have
used the version {v9r32p5} of the {\it Fermi}-LAT ScienceTools\footnote{http://fermi.gsfc.nasa.gov/ssc/data/analysis/scitools/overview.html} {and the P7REP\_SOURCE\_V15 IRF}.
From the downloaded data we have discarded events taken at zenith angles
above 100$\degr$  to reduce the contamination of albedo {gamma rays} coming from the Earth's
limb. To compute the pulse phase, we used the same spin parameters as for the MAGIC analysis. The obtained fluxes were
computed by maximizing the likelihood of a given source model using the gtlike tools. {The binned likelihood method was adopted and a 40\degr~square area with 0.2\degr~bin width was used for the likelihood maximization.} Apart from the Galactic {({\sf gal\_iem\_v05.fits})} and extragalactic {({\sf iso\_source\_v05.txt})} diffuse emission, we considered as background sources for the likelihood fits {all sources listed} in the second LAT source catalogue \citep{2012ApJS..199...31N}. {The data taken during the periods when the Crab nebula was flaring were not excluded from the analysis. These flares should not have any impact on the pulsed emission results because it is known that the pulsation component did not change during the flares \citep{2012ApJ...749...26B}, and the average nebula flux including flare periods was subtracted when the pulsar signal was determined. Regarding the reported Fermi-LAT spectrum from the Crab nebula, the six Crab flares that lasted a few days might be responsible for a few percent of the photons below 1~GeV in the overall 5.5 year dataset. Given that the effect is expected to be small, and that this paper focusses on the emission from the pulsar, we did not correct for this effect.}


\section{Results}

\subsection{Light curve above 50 GeV}
\label{SectMAGICLC}
\begin{figure*}
\centering
\includegraphics[width=13cm]{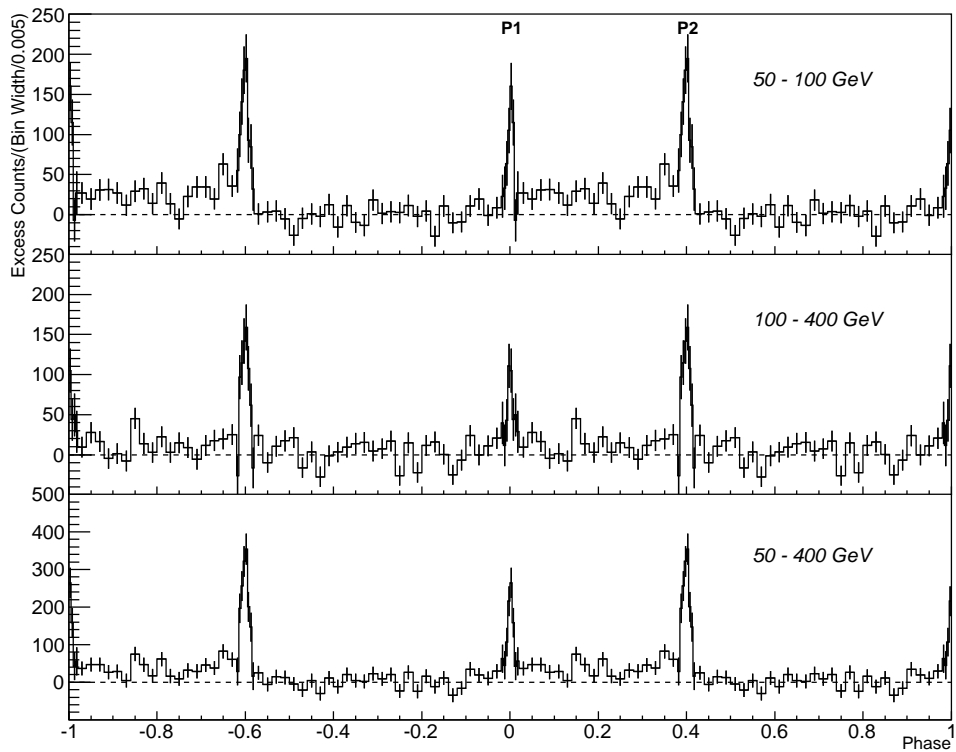} 
\caption{Light curves of the Crab pulsar obtained by MAGIC from 50~GeV to 100~GeV (top), from 100~GeV to 400~GeV (middle), and for the full analyzed energy range (bottom). The bin widths around the peaks are 4 times smaller {(0.005)} than the rest {(0.02)} in order to highlight the sharpness of the peaks.}
\label{RefFig1}
\end{figure*}

Figure \ref{RefFig1} shows the light curves of the Crab pulsar measured by MAGIC.
Two peaks are clearly visible.
Following our previous study \citep{2012A&A...540A..69A}, we define phase ranges for the two peaks as
P1$_{\rm M}$ (phase $-0.017$ to $0.026$) and P2$_{\rm M}$ (0.377 to 0.422). The background level (hadrons and continuum {gamma rays}) is estimated using the phase range between 0.52 and 0.87 {and it is then subtracted from the histograms}\footnote{
An estimation of the background using the off-peak interval from the LAT Second Pulsar Catalog, namely the phase range between 0.61 and 0.89, lead to very similar results.}.
The number of excess events in P1$_{\rm M}$ between 50~GeV and 400~GeV is $930 \pm 120$ ($8.0~\sigma$) and in P2$_{\rm M}$ is $1510  \pm 120$ ($12.6 ~\sigma$). 

In addition to the two main peaks, significant emission between them is also visible. 
The region between the peaks is generally called the Bridge.
Defining {the} Bridge region as the gap between P1$_{\rm M}$ and P2$_{\rm M}$, namely, between 0.026 and 0.377 (hereafter Bridge$_{\rm M}$), 
we obtain an excess of $2720 \pm 440$ (6.2~$\sigma$) events in this region.
Adopting the definition used at lower energies for the Bridge as the region $0.14 - 0.25$ from \citet{1998ApJ...494..734F} 
 (hereafter Bridge$_{\rm E}$),  then the number of excess events 
is $880 \pm 200$ (4.4 $\sigma$). This excess increases to  
 $1940 \pm 370$ (5.2 $\sigma$) if we extend Bridge$_{\rm E}$ with the so-called trailing wing of P1 and the leading wing of P2, namely to the interval of $0.04 - 0.32$ \citep[see][]{1998ApJ...494..734F}. {It should be noted that this detection confirms the hint of bridge emission already reported in \citep{2012A&A...540A..69A}}.

\subsection{Comparison with lower energies}

\begin{figure}
\centering
\resizebox{9cm}{!}{ \includegraphics[width=3cm]{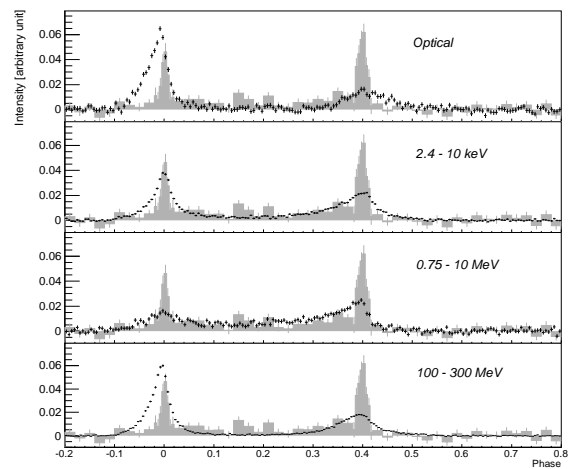} }
\caption{
Light curve of the Crab pulsar at optical wavelength, $2.4-10$~keV X-rays, $0.75-10$~MeV, and $100-300$~MeV gamma rays (from top to bottom). The light curve at $50-400$ GeV is overlaid on each plot for comparison. The optical light curve was obtained with the MAGIC telescope using the central pixel of the camera \citep{2008NIMPA.589..415L}. The keV and MeV light curves are from \cite{2001A&A...378..918K}. The $100-300$~MeV light curve was produced using the {\it Fermi}-LAT data. {All light curves are zero-suppressed by estimating the background using the events in the phase range from 0.52 to 0.87.} }
\label{RefFig2}
\end{figure}

Figure~\ref{RefFig2} shows the light curves at optical, X-ray, and gamma-ray energies obtained with various instruments, together with the $50-400$~GeV light curve from the bottom panel of Fig.~\ref{RefFig1}. {The background was subtracted in the same way as the MAGIC light curves (see Sect.~\ref{SectMAGICLC})}.
 The intensity and morphology of the bridge emission varies considerably with energy. It is very weak at optical wavelengths and in the $100-300$ MeV range, while there is an appreciable difference at X-rays and soft {gamma rays}. At the energies covered by MAGIC, the peaks get much sharper and a prominent bridge emission appears.


\begin{figure}
\centering
\resizebox{9.0cm}{!}{ \includegraphics[width=3cm]{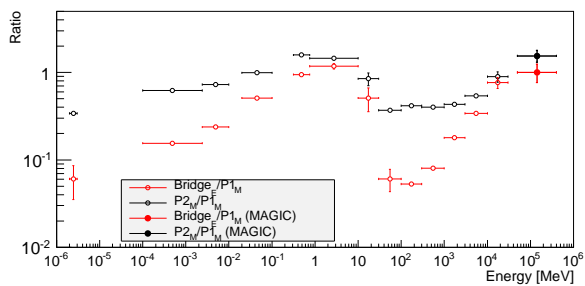} }
\caption{ P2$_{\rm M}$/P1$_{\rm M}$ ratio (black markers) and Bridge$_{\rm E}$/P1$_{\rm M}$ ratio (red markers) as a function of energy.
At optical energies (a few eV), the ratios are obtained using the central pixel of the MAGIC camera \citep{2008NIMPA.589..415L}. From 100 eV to 100 MeV, ratios are computed based on the light curves shown in \citet{2001A&A...378..918K}.
From 100 MeV to 30 GeV, {\it Fermi}-LAT data were used.}
\label{RefFig3}
\end{figure}



It is known that the flux ratio between the two peaks strongly depends on energy, as does the ratio between the first peak and the bridge \citep[see, e.g.,][]{2001A&A...378..918K}.
Fig. \ref{RefFig3} shows the flux ratio between P2$_{\rm M}$  and P1$_{\rm M}$  and that between Bridge$_{\rm E}$ and P1$_{\rm M}$ as a function of energy from optical ($\sim 2$ eV) to 400 GeV. {Steady emission was subtracted before the ratios were computed}.  
The ratios P2$_{\rm M}$/P1$_{\rm M}$ and Bridge$_{\rm E}$/P1$_{\rm M}$ behave similarly. These ratios increase with energy up to 1 MeV, decrease up to 100 MeV, and increase again from that energy on. 
At $50 - 400$ GeV, the ratios basically follow the trend seen at lower energies.


\subsection{Spectral energy distribution}
\begin{figure}
\centering
\resizebox{9cm}{!}{
 \includegraphics[width=6cm]{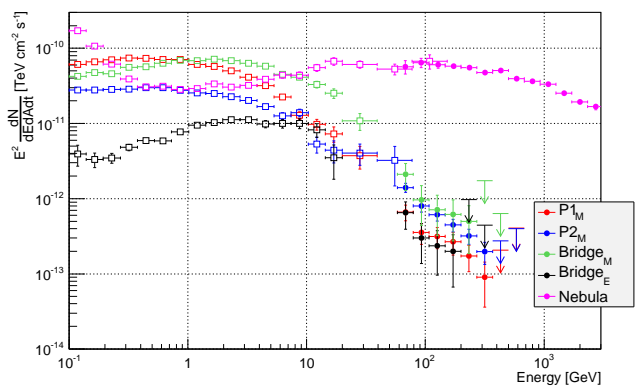} 
 }
\caption{{Spectral energy distributions} of the Crab nebula, P1$_{\rm M}$, P2$_{\rm M}$, Bridge$_{\rm M}$, and Bridge$_{\rm E}$ measured with {\it Fermi}-LAT (below 50 GeV) and MAGIC (above 50 GeV). {The flux values averaged over the rotation period are plotted.} }
\label{RefFig4}
\end{figure}

The spectral energy distributions (SEDs) of the P1$_{\rm M}$, P2$_{\rm M}$, Bridge$_{\rm M}$, and Bridge$_{\rm E}$ between 100 MeV and 400 GeV are shown in Fig. \ref{RefFig4}, together with the Crab nebula SED obtained with a subset of the data used for the pulsar analysis. 
The SEDs were calculated using {\it Fermi}-LAT data below 50 GeV (below 200 GeV for the nebula), and MAGIC data above 50 GeV. The nebula SED is connected smoothly between the two instruments.
The {\it Fermi}-LAT data were fit with a power law with an exponential cutoff, while the MAGIC data were fit with a simple power{-}law function.
The obtained fit parameters are summarized in Table~\ref{RefTab1}. The power{-}law indices between 50~GeV and 400~GeV are about 3 and no significant difference is seen between different pulse phases. The uncertainty in the absolute energy scale is estimated  as ~17\%, whereas the systematic error of the flux normalization is estimated to be ~18\%. The difference between this number and the one given in \cite{2012APh....35..435A} is mainly due to a more precise background estimation from the off-peak region. We estimate the overall systematic uncertainty uncertainty on the spectral slope to be 0.3.

\begin{table*}
\caption{Spectral Parameters}
\label{RefTab1}
\begin{tabular}{c|c|c|c|c|c}
\hline
\hline
phase & $F_1$\tablefootmark{a}  {\tiny [$10^{-11} {\rm MeV}^{-1} {\rm cm}^{-2} {\rm s}^{-1}]$}  & $\Gamma_1$\tablefootmark{a} & $E_c$\tablefootmark{a} {\tiny [GeV]}& $F_{100} \tablefootmark{b}$ {\tiny [$10^{-11} {\rm TeV}^{-1} {\rm cm}^{-2} {\rm s}^{-1}]$ }& $\Gamma_2$\tablefootmark{b}\\
\hline
P1$_{\rm M}$ & $8.87 \pm 0.14$ & $1.88 \pm 0.01$ & $3.74 \pm 0.15$ &$4.18 \pm 0.59$ & $3.25 \pm  0.39 $ \\
P2$_{\rm M}$ & $3.14 \pm 0.07$ & $1.97 \pm 0.01$ & $ 7.24 \pm 0.64$& $8.48 \pm 0.62$ & $3.27 \pm 0.23$ \\
Bridge$_{\rm M}$ & $7.70 \pm 0.11$& $1.74 \pm 0.01$ & $7.19 \pm 0.39$ &$ 12.2 \pm 3.3$ & $ 3.35 \pm 0.79$\\
Bridge$_{\rm E}$ & $0.95 \pm 0.04$& $1.44 \pm 0.04$ & $6.94 \pm 0.90$ &$ 3.7 \pm 1.1$ & $ 3.51 \pm 0.97$\\
\hline
\end{tabular}
\\
\tablefoottext{a}{{\tiny Spectral parameters obtained by fitting a function $F(E)~=~F_1 (E/1{\rm GeV})^{-\Gamma_1} \exp(E/E_{c})$ to {\it Fermi}-LAT data between 100 MeV and 300 GeV}}\\
\tablefoottext{b}{{\tiny Spectral parameters obtained by fitting a function $F(E)~=~F_{100} (E/100{\rm GeV})^{-\Gamma_2}$ to MAGIC data between 50 GeV and 400 GeV}}
\end{table*}

\section{Discussion}
\input{Crab_Bridge_hirotani4}

\begin{acknowledgements}
We would like to thank Felix Aharonian and Dmitry Khangulyan for useful discussions on the
pulsar emission models. 
We would also like to thank the Instituto de Astrofísica de Canarias for
the excellent working conditions at the Observatorio del Roque de los Muchachos in La Palma.
The support of the German BMBF and MPG, the Italian INFN, the Swiss National Fund SNF,
and the Spanish MINECO is gratefully acknowledged. This work was also supported by the
CPAN CSD2007-00042 and MultiDark CSD2009-00064 projects of the Spanish Consolider-
Ingenio 2010 programme, by grant DO02-353 of the Bulgarian NSF, by grant 127740 of the
Academy of Finland, by the DFG Cluster of Excellence “Origin and Structure of the Universe”,
by the DFG Collaborative Research Centers SFB823/C4 and SFB876/C3,  by the Polish
MNiSzW grant 745/N-HESS-MAGIC/2010/0, by  the Croatian Science Foundation (HrZZ) Project 09/186 and by the Formosa Program between National Science Council 
in Taiwan and Consejo Superior de Investigaciones Cientificas in Spain administered through grant number NSC100-2923-M-007-001-MY3.

\end{acknowledgements}

\bibliographystyle{bibtex/aa}
\bibliography{EnglishCorrection}

\end{document}

%% file: authors_last_AA_140210.tex
%
\author{
J.~Aleksi\'c\inst{1} \and
S.~Ansoldi\inst{2} \and
L.~A.~Antonelli\inst{3} \and
P.~Antoranz\inst{4} \and
A.~Babic\inst{5} \and
P.~Bangale\inst{6} \and
U.~Barres de Almeida\inst{6} \and
J.~A.~Barrio\inst{7} \and
J.~Becerra Gonz\'alez\inst{8,}\inst{26} \and
W.~Bednarek\inst{9}
% ...
}
\institute { IFAE, Campus UAB, E-08193 Bellaterra, Spain
\and Universit\`a di Udine, and INFN Trieste, I-33100 Udine, Italy
\and INAF National Institute for Astrophysics, I-00136 Rome, Italy
\and Universit\`a  di Siena, and INFN Pisa, I-53100 Siena, Italy
\and Croatian MAGIC Consortium, Rudjer Boskovic Institute, University of Rijeka and University of Split, HR-10000 Zagreb, Croatia
\and Max-Planck-Institut f\"ur Physik, D-80805 M\"unchen, Germany
\and Universidad Complutense, E-28040 Madrid, Spain
\and Inst. de Astrof\'isica de Canarias, E-38200 La Laguna, Tenerife, Spain
\and University of \L\'od\'z, PL-90236 Lodz, Poland
\and Deutsches Elektronen-Synchrotron (DESY), D-15738 Zeuthen, Germany
\and ETH Zurich, CH-8093 Zurich, Switzerland
\and Universit\"at W\"urzburg, D-97074 W\"urzburg, Germany
\and Centro de Investigaciones Energ\'eticas, Medioambientales y Tecnol\'ogicas, E-28040 Madrid, Spain
\and Technische Universit\"at Dortmund, D-44221 Dortmund, Germany
\and Universit\`a di Padova and INFN, I-35131 Padova, Italy
\and Universit\`a dell'Insubria and INFN Milano Bicocca, Como, I-22100 Como, Italy
\and Unitat de F\'isica de les Radiacions, Departament de F\'isica, and CERES-IEEC, Universitat Aut\`onoma de Barcelona, E-08193 Bellaterra, Spain
\and Institut de Ci\`encies de l'Espai (IEEC-CSIC), E-08193 Bellaterra, Spain
\and Japanese MAGIC Consortium, Division of Physics and Astronomy, Kyoto University, Japan
\and Academia Sinica, Institute of Astronomy and Astrophysics (ASIAA), P.O. Box: 23-141, Taipei, Taiwan
\and Finnish MAGIC Consortium, Tuorla Observatory, University of Turku and Department of Physics, University of Oulu, Finland
\and Inst. for Nucl. Research and Nucl. Energy, BG-1784 Sofia, Bulgaria
\and Universitat de Barcelona, ICC, IEEC-UB, E-08028 Barcelona, Spain
\and Universit\`a di Pisa, and INFN Pisa, I-56126 Pisa, Italy
\and ICREA and Institut de Ci\`encies de l'Espai (IEEC-CSIC), E-08193 Bellaterra, Spain
\and now at: NASA Goddard Space Flight Center, Greenbelt, MD 20771, USA and Department of Physics and Department of Astronomy, University of Maryland, College Park, MD 20742, USA
% ...
}

%% file: Crab_Bridge_hirotani4.tex
In summary, the Crab pulsar above 50 GeV 
exhibits a light curve with a significant bridge emission between two sharp peaks (Fig.~1).  
The flux ratios P2$_{\rm M}$/P1$_{\rm M}$ and Bridge$_{\rm E}$/P1$_{\rm M}$ increase with increasing photon 
energy between 100~MeV and 400~GeV (Figs.~2 and 3). 
Between 30~GeV and 400~GeV,
the fluence in the bridge phase is comparable to that in the P1 phase (Fig.~4). 
The SEDs in the $50-400$~GeV range could be fit with power-law functions for the three phases.

Detection of pulsed VHE emissions favors emission sites in the outer 
part of the magnetosphere 
because a strong source attenuation is expected at lower altitudes at these energies.
The outer-gap (OG) and the slot-gap models are the most 
probable explanation of these pulsed $\gamma$-rays 
\citep{2008ApJ...680.1378H,2011ApJ...727..123W,2012ApJ...744...34V}.
Using an ad hoc extension of the two dimensional meridional OG model to three dimension, 
\cite{2008ApJ...676..562T} and \cite{2008MNRAS.386..748T} 
reproduced the bridge emission.
However, a fully three-dimensional electrodynamical structure is required to model the phase resolved SEDs \citep{2011ApJ...733L..49H,2013ApJ...766...98H}.

Alternatively, if a very strong magnetic-field-aligned electric field arises near the light cylinder (LC), pulsed VHE photons might be also emitted there \citep{2012MNRAS.424.2079B}.
Emission from beyond the LC can also explain the double-peaked light curves.
\cite{2013A&A...550A.101A}
demonstrated that a sufficient luminosity and a hard spectrum extending 
to 100~GeV can be obtained for P1 and P2 via the synchrotron emission 
by a hot plasma from the current sheet slightly outside the LC, but in this scenario the bridge emission 
should disappear above 10~GeV. 
\cite{2013ApJ...773..143C}
proposed that synchrotron radiation generated near 
the LC during the quasi-linear stage of the cyclotron instability 
can produce the phase-aligned 
pulsation between radio and $\gamma$-rays. 
However, the formation of a bridge component is not explained
in this model.  

Although synchrotron luminosity declines sharply beyond the LC, 
the inverse-Compton process may still be effective there.
\cite{2012Natur.482..507A}
demonstrated that the observed pulsed flux of the Crab pulsar between 70~GeV 
and 400~GeV can be explained by up-scattered photons by a particle-dominated wind
whose Lorentz factors exceed $5 \times 10^5$ at $20-50$ LC radii. 
Although a phase-resolved spectrum is not provided in their paper,
the observed P2/P1 ratio in VHE could be reproduced 
if one considers an anisotropic wind. 
{The bridge emission is also predicted, but a special density profile is required to explain both the bridge and the narrow peak emissions at the same time \citep{2012AIPC.1505...29K}.}

In closing, none of the current models can consistently account for the properties of the pulsed and bridge emission from the Crab pulsar.
  













